\documentclass{aastex63}
\usepackage{hyperref}
\usepackage{booktabs}
\submitjournal{ApJ}

\shorttitle{Low-Latitude Auroras: Unravelling the Mystery}
\shortauthors{Vichare et al.}
\graphicspath{{./}{figures/}}

\begin{document}

\title{Low-Latitude Auroras: Insights from 23 April 2023 Solar Storm}

\correspondingauthor{Geeta Vichare}
\email{geeta.vichare@iigm.res.in}

\author [0000-0003-3607-6923]{Geeta Vichare}
\affiliation{Indian Institute of Geomagnetism, Navi Mumbai, Maharashtra, India.
}

\author[0000-0003-4281-1744]{Ankush Bhaskar}
\affiliation{Space Physics Laboratory, ISRO/Vikram Sarabhai Space Centre, Thiruvananthapuram, India}

\author{ Rahul Rawat   }
\affiliation{Indian Institute of Geomagnetism, Navi Mumbai, Maharashtra, India.
}

\author{Virendra Yadav}
\affiliation{Aryabhatta Research Institute of Observational Sciences, Nainital, Uttarakhand, India.
}

\author{Wageesh Mishra }
\affiliation{Indian Institute of Astrophysics, Bengaluru, Karnataka, India.
}

\author{ Dorje Angchuk}
\affiliation{Indian Institute of Astrophysics, Bengaluru, Karnataka, India.
}

\author{ Anand Kumar Singh}
\affiliation{National Centre for Polar and Oceanic Research, Goa, India}



\begin{abstract}

In April 2023, low-latitude aurora observation by the all-sky camera at Hanle, Ladakh, India ($33^{\circ} {} N $ geographic latitude (GGLat)) was reported, which stimulated a lot of discussion among scientists as well as masses across the globe. The reported observation was intriguing as the solar storm that triggered this aurora was moderate and the first such observation from Indian region in the space-era. In this communication, we investigate such a unique modern-day observation of low-latitude auroral sighting occurring during the passage of sheath-region of Interplanetary-Coronal-Mass-Ejection, utilizing in situ multi-spacecraft particle measurements along with geomagnetic-field observations by ground and satellite-based magnetometers. Auroral observations at Hanle coincided with the intense substorm occurrences. It is unequivocally found that the aurora didn’t reach India, rather the equatorward boundary of the aurora was beyond $ 50^{\circ} {}N $ GGLat. The multi-instrumental observations enabled us to estimate the altitude of the red auroral emissions accurately. The increased flux of low-energy electrons ($<$100 eV) precipitating at $\sim 54^{\circ}N$ GGLat causing red-light emissions at higher altitudes ($\sim$700-950 km) can be visible from Hanle. The observed low-latitude red aurora from India resulted from two factors: emissions at higher altitudes in the auroral oval and a slight expansion of the auroral oval towards the equator. The precipitating low-energy particles responsible for red auroral emissions mostly originate from the plasma sheet. These particles precipitate due to wave-particle interactions enhanced by strong compression of the magnetosphere during high solar wind pressure. This study using multi-point observations holds immense importance in providing a better understanding of low-latitude auroras.

\end{abstract}


\section{Introduction} \label{sec:intro}
The plasma particles and electromagnetic fields of solar origin interact with the Earth’s magnetic field, leading to several space weather phenomena such as geomagnetic storms, substorms, and visible aurora \citep{gonzalez1994geomagnetic,kilpua2015statistical, balan2019capability,zhao2022can, raghav2018torsional,dai2023geoeffectiveness}. There have been several solar storms impinging the Earth and resulting in geomagnetic storms and even low-latitude aurora \citep{hajra2018interplanetary, hayakawa2018low,hayakawa2023extreme,wang2021aurora,hayakawa2022temporal}. The solar energy injected into the Earth’s magnetosphere through coupling gets distributed in various regions of the magnetosphere \citep{vichare2005some}. The substorms are electrodynamic events that occur in the Earth's magnetosphere-ionosphere system, characterized by intense auroral displays in the latitudinal ring between $\sim 60^\circ$ and $75^\circ$ in both the hemispheres, also known as ‘auroral oval’ \citep{akasofu1964development, rostoker1980magnetospheric, baker1996neutral}. The Aurora Borealis (Northern Lights) is visible in countries like Norway, Finland, Iceland, Russia, Alaska, Greenland, etc., and Aurora Australis  (Southern Lights) is mainly observed in Antarctica, and sometimes in Australia. When fluxes of energetic electrons and protons precipitate along the magnetic field lines to enter into the Earth’s atmosphere, auroral emissions occur. An increase in the convective electric field produces intense currents in the ionosphere, giving rise to auroral electrojets, monitored through magnetic field measurements  \citep{mcpherron1972substorm,newell2011substorm,behera2015substorm}. 
Isolated Proton Aurora (IPA), Strong Thermal Emission Velocity Enhancement (STEVE), and Stable Auroral Red (SAR) are the most prominent optical emissions in the sub-auroral regions \citep{gallardo2021proton, nishimura2022interaction}. IPA is found to be caused by fine-structured EMIC Pc1 waves \citep{sakaguchi2015isolated,nomura2016pulsating,kim2021isolated}. SAR arcs are considered to be due to energy transfer from the ring current to the hot plasma in the outer plasmasphere \citep{cole1965stable,cole1970magnetospheric}. Whereas, STEVE is a heated glowing gas created by very fast plasma streams and appears as streaks in the sky during auroral activity. Thus at the equatorward boundary of the auroral oval, a variety of optical emissions with different causes exist. \citet{shiokawa1997multievent} showed that the precipitation of broadband electrons (BBE) is responsible for the red aurora at mid-latitudes. On examining long historical records, \citet{lee2023response} found that red auroras frequently appear in low/middle latitudes during high solar activity periods. Further, instances of mid to low-latitude aurorae have been documented during periods of relatively calm to moderately active geomagnetic conditions as well. \citet{silverman2003sporadic} categorized these aurorae as  ``Sporadic Aurorae".  Past research has identified sporadic aurorae in mid- to low-latitude regions (e.g. \citet{hayakawa2018sporadic,oliveira2020possible,vaquero2007sporadic, vaquero2013possible, willis2007sporadic, wang2021aurora}). These aurorae are typically faint, and their specific characteristics, including duration and their association with compression of the magnetosphere and substorms, are subjects of ongoing investigation \citep{silverman2003sporadic,hayakawa2018sporadic, oliveira2020possible, bhaskar2020analysis}. Understanding this phenomenon poses a challenge in the context of magnetospheric physics. Maybe one needs to unravel the mechanisms through which a bunch of energetic particles can be transported deep into the closed field lines of the magnetosphere. While the precise mechanisms behind these aurorae remain incompletely understood, it is of paramount importance to observe such events during the modern era when a number of state-of-the-art space and ground-based observing facilities are available.  While most of the sporadic or low-latitude auroral observations are prior to the space age and inferred from historical records, the present study provides an opportunity to investigate the low-latitude auroral observations for one of the recent moderate-level geomagnetic storms.

\begin{figure}[h!]
	\centering
	\includegraphics[width=15.0cm]{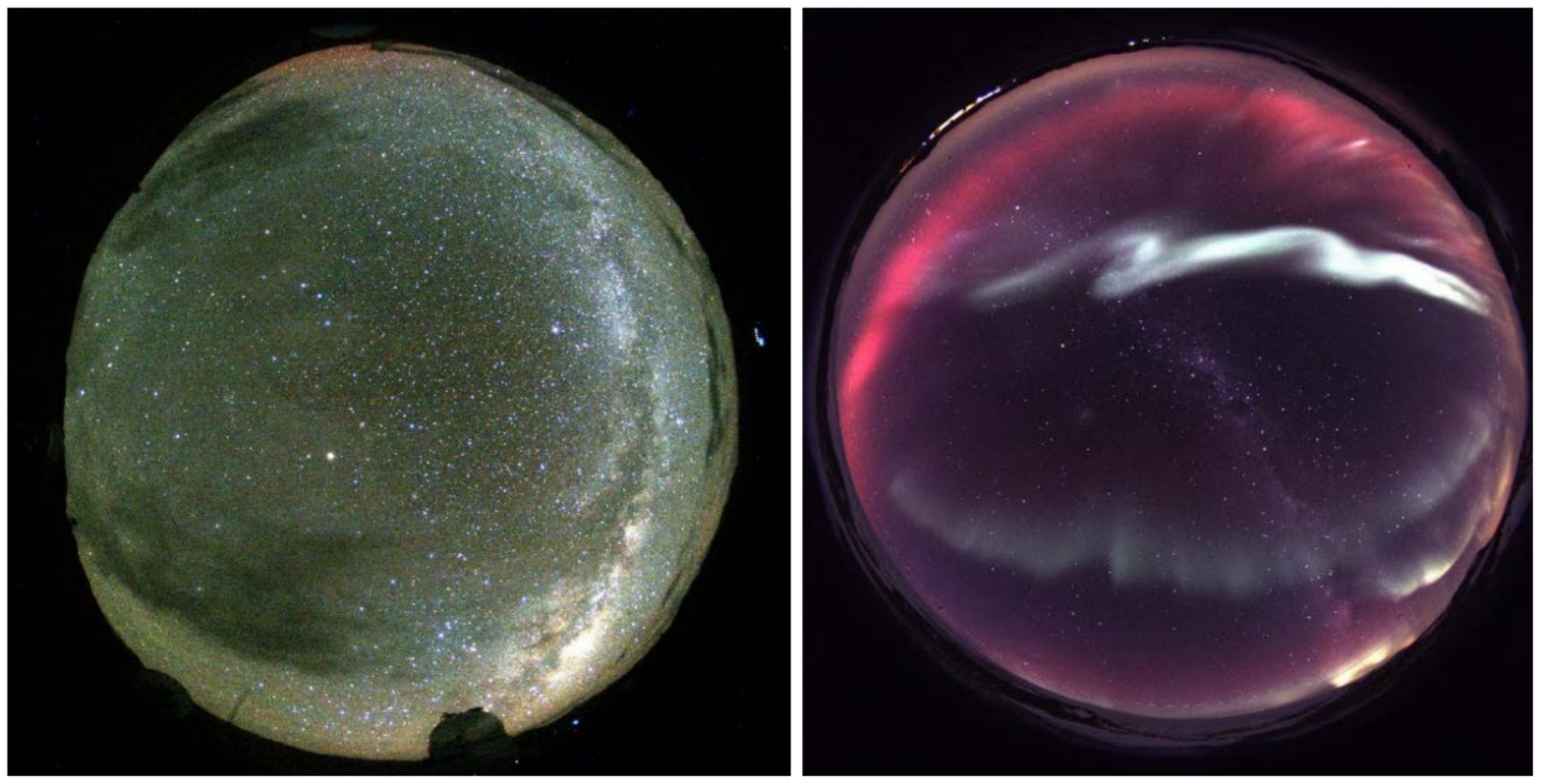}
	\caption{Aurora seen on 23 2023 by all-sky camera from (Left) Hanle, Ladakh, India at 19:54 UT, (Right) Indian Antarctic station, Maitri at 19:27 UT. The north direction is Upward. The red diffuse glow is visible on the top side in the Left panel.
}
	\label{fig:aurora_photo}
\end{figure}

On the night of 23-24 April 2023, an aurora was observed using an all-sky camera ($180^\circ$ field of view) at Hanle in Ladakh, India. The reported observation was intriguing as the geomagnetic storm during this observation was moderate and not very intense. Hence it was puzzling to understand how it was possible to observe the aurora at such a low latitude. The first task was to confirm that the observed red emission at the north of the Hanle is not an airglow, as aurora and airglow both are naturally occurring luminescent phenomena in the Earth’s atmosphere. However, there is a basic difference between these two i.e. the aurora appears due to the interaction between energetic particles and the ionosphere, while airglow is caused by solar radiation. Airglow is relatively fainter and occurs globally and at all times, whereas aurora appears during substorms. Hanle observation sources reported that the observed red emission is not an airglow and it appears to be a ``stable auroral red (SAR) arc”. This is the first all-sky camera record of low-latitude aurora from India, unlike the past historical records of the phenomenon through human naked-eye observations.  In this communication, we investigate this event in detail using satellite-based measurements of particle flux and field-aligned currents, and ground-based magnetic field observations to unravel the mystery of auroral observations at low latitudes.

\section{Hanle Observations} 

Following the Indian Institute of Astrophysics' report on the Aurora at Ladakh, false pictures claiming to be the recent Aurora in Ladakh were spread on social media. However, those were old pictures taken in Norway and Iceland, and not in Ladakh.
Figure \ref{fig:aurora_photo} (left side panel) shows the actual picture of the aurora (red glow at the top side of image) seen by the All Sky camera at Hanle at 1:24 IST on 24 April (19:54 UT on 23 April), and right side panel shows that at Indian Antarctic station, Maitri at 19:27 UT. The red light that appeared on the north of Hanle has a zenith angle of $\sim 80^\circ - 85^\circ  N $. This indicates that the aurora has not occurred over Hanle but at higher latitudes. The red emission was observed from 19:30 UT (01:00 IST) till 22:00 UT (03:30 IST), and later on it became very faint. At 22:30 UT (04:00 IST) it appeared pinkish. It may be noted that the observations from the all-sky camera at Hanle are possible during the dark hours only and hence observations during sunlit hours are not available.  The elevation angle of the red emission at Hanle seen in Figure \ref{fig:aurora_photo} is computed using background star positions and a simulated view of the night sky at Hanle using Stellarium planetarium software. The elevation angle of the top edge of the red glow was found to be around  $6^\circ$ to $8^\circ$. Note that, the auroral image of Antarctica also shows the occurrence of red emissions.

\section{Interplanetary and geomagnetic conditions}

\begin{figure}[h!]
	\centering
	\includegraphics[width=12.0cm]{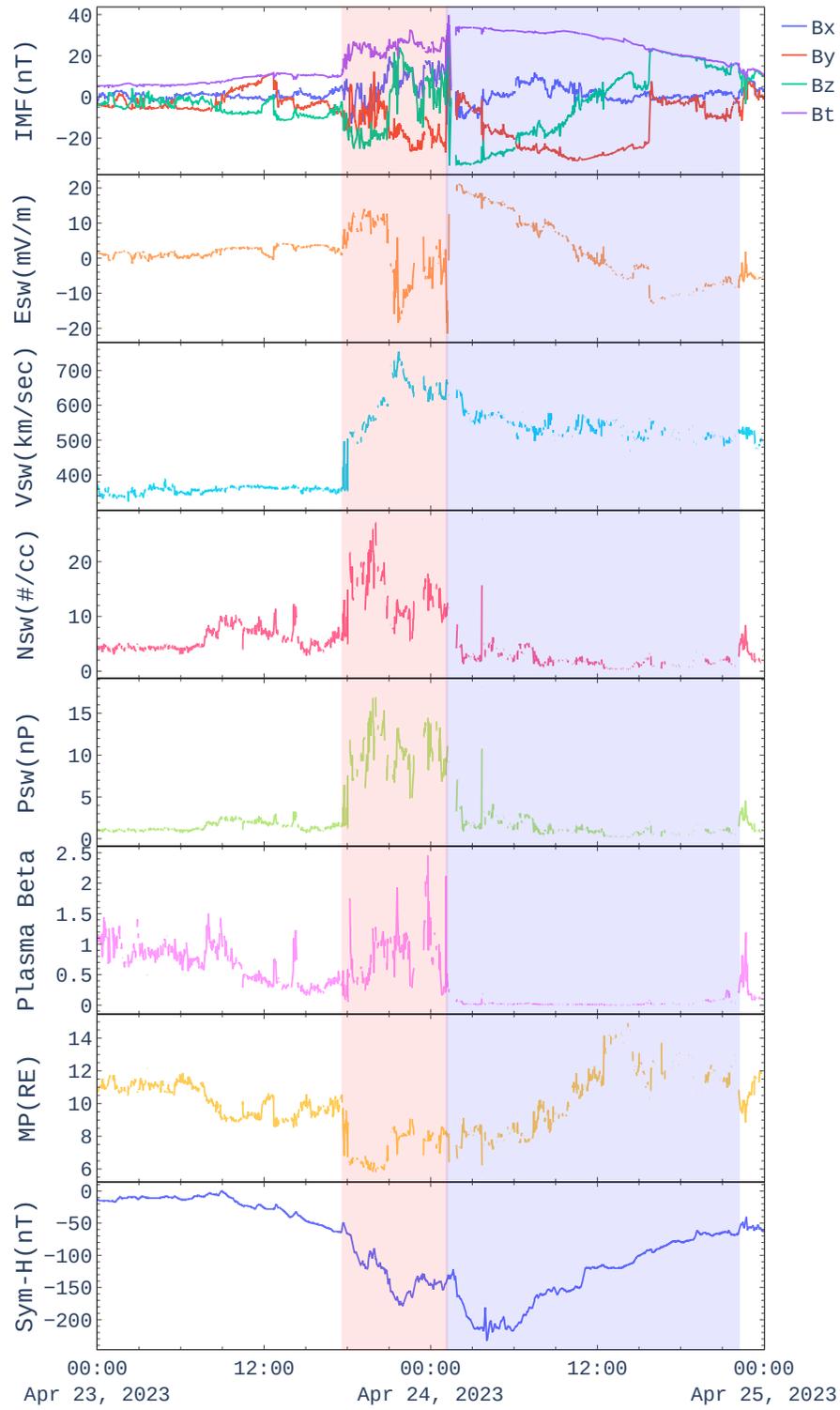}
	\caption{  Interplanetary and solar wind parameters along with magnetopause distance (MP) and geomagnetic activity index SYM-H during 23-24 April 2023. The shaded region in pink indicates sheath duration whereas in violet shade magnetic cloud is shown.}
	\label{fig:IP}
\end{figure}
On 21 April 2023, M1.7 class solar flare along with a coronal mass ejection (CME) was ejected from the Sun, causing a moderate geomagnetic storm on 23-24 April on the Earth \citep{vemareddy2024filament}. Figure \ref{fig:IP} shows the interplanetary parameters and Sym-H index during 23-24 April 2023. After 8 UT on 23 April, the north-south component (Bz) of the interplanetary magnetic field (IMF) turned southward and remained southward for several hours facilitating the magnetic field reconnection at the Earth’s magnetopause. However, a sharp southward turning of IMF took place just before 18 UT, which resulted in an enhanced geomagnetic disturbance on Earth. The sheath region showed elevated plasma parameters and some quasi-planarity structure \citep{ghag2024quasi}. This disturbance ($SymH \sim -175$ nT) on 23 April is mainly caused during the passage of the sheath region of Interplanetary CME (ICME), whereas further enhancement of the ring current with $SymH= -200$ nT on 24 April is caused by the magnetic cloud of the ICME. The geomagnetic storm monitored by Sym-H index indicates that the ring current started slowly developing from 9 UT on 23 April, as the IMF Bz component was almost -10 nT, and then before 18 UT, the ring current strengthened due to further southward turning of the IMF, with the corresponding eastward component of interplanetary electric field (Ey) reaching $\sim 15mV/m$. This is also accompanied by a sudden increase in the solar wind dynamic pressure (Psw), resulting in sudden commencement at 17:35 UT. At 21 UT, IMF turned northward. The geomagnetic storm index Sym-H attained a value of -175 nT at $\sim 22$ UT. Later, on 24th April, Sym-H became stronger crossing -200 nT. The plasma beta and temperature parameters indicate that the geomagnetic storm has taken place during the magnetic cloud of the CME.

\begin{figure}[h!]
	\centering
	\includegraphics[width=15.0cm]{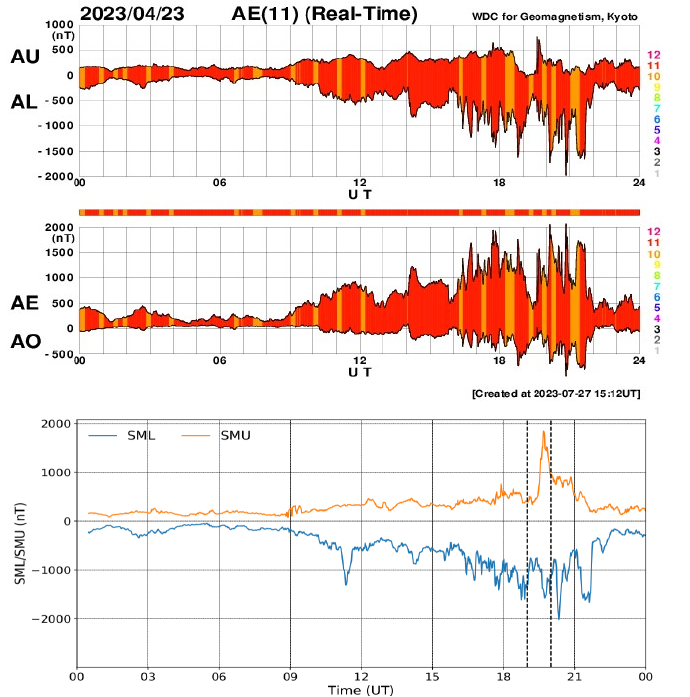}
	\caption{ (Top two panels) Preliminary Auroral electrojet indices on 23rd April 2023 from WDC, Kyoto. (Bottom panel) SMU and SML indices representation auroral activity obtained from Supermag (\url{https://supermag.jhuapl.edu/})}
	\label{fig:AL_AU_kyoto}
\end{figure}

Figure \ref{fig:AL_AU_kyoto} depicts the AL, AU, AE, and AO indices (preliminary) obtained from WDC Kyoto and SML, SMU indices from Supermag. The geomagnetic disturbance in the H-component, as observed at 10-12 longitudinally distributed observatories located in the auroral zone of the northern hemisphere, is used to derive the auroral electrojet indices \citep{davis1966auroral}. The sudden decrease in the AL index indicates the start of the substorm expansion phase and defines the substorm onset time. From Figure  \ref{fig:AL_AU_kyoto}, it can be observed that on 23 April, the substorm activity was significantly enhanced after 10 UT. It can be noticed that there were episodes of super substorms with AL/SML reaching -1500 nT during 16-22 UT ($21:30$ to $3:30$ IST). Thus, almost the entire night over Indian longitudes, there was the occurrence of substorms.

\section{Magnetic field variations}

\begin{figure}[h!]
	\centering
	\includegraphics[width=15.0cm]{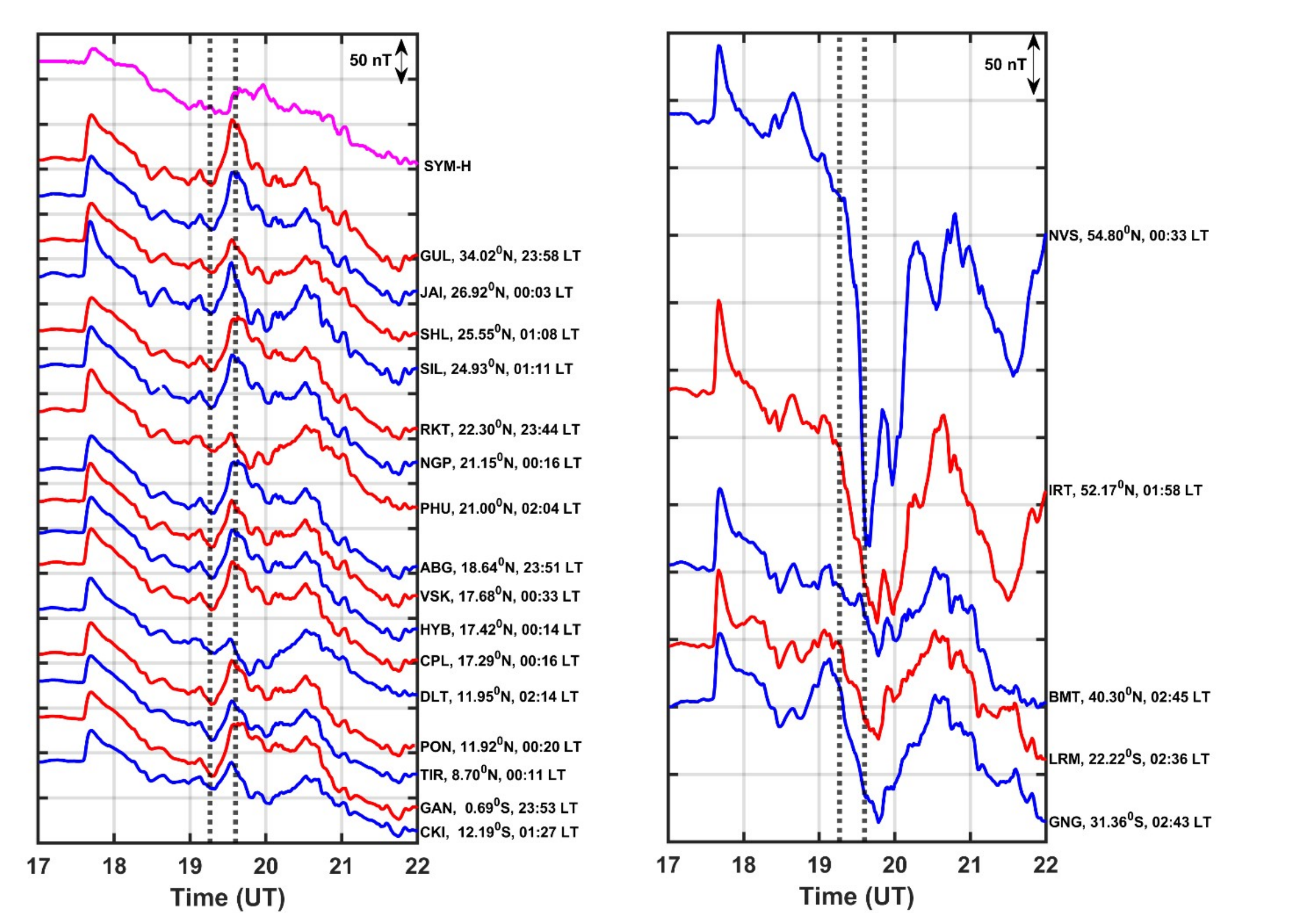}
	\caption{ Horizontal magnetic field variation at different latitudes between $70^\circ-115^\circ$E longitude sectors. SymH index is shown on top in the left side panel. Station code, geographic latitude, and LT at the first dotted vertical line are shown on the right side of each plot. Left and right panels are separated based on the nature of variation between two vertical dotted lines.
}
	\label{fig:mag_var}
\end{figure}

In addition to high latitudes, mid-latitude magnetic field measurements are also used to study the substorm-associated current system. Substorm current wedge (SCW) produces systematic positive H variations at mid-latitudes (also called positive bay) on the night side, particularly between 21-02 LT hours, at the time of the substorm expansion phase \citep{mcpherron1973satellite,clauer1974mapping, dejong2007aurora}. Figure \ref{fig:mag_var} shows the variation of the H component at the observatories from the Eastern Asian region ($70^\circ-115^\circ$ E longitude), during 17-22 UT on 23 April 2023. This time window represents the local night in this sector. Table 1 shows the list of observatories used along with the geographic (GG), geomagnetic (GM), and corrected geomagnetic (CGM) coordinates. Figure \ref{fig:mag_var} displays the geomagnetic variations from mid-latitude stations (alternate red and blue curves) in two panels. For comparison purposes, the Sym-H index is plotted topmost on the left panel (magenta color). As discussed in section 3, ICME shock at $\sim$ 17:35 UT has resulted in a sudden increase in the H component on the ground, which is also reflected in the Sym-H index. This was followed by the ring current development, which is seen as a declining H component at almost all mid-latitude stations. Though several episodes of substorm activity occurred during this time, the mid-latitude variations are complex due to ongoing geomagnetic storm, and hence difficult to identify substorm-associated positive bay. The figure presented in the supplementary material (Figure S1) obtained by applying a band pass filter between 5-25 mHz, shows the presence of Pi2 pulsations during 19-21 UT, confirming the substorm activity. The H component between the vertical dashed lines in Figure \ref{fig:mag_var} depicts positive variations at the stations shown on the left panel and negative variations at the stations on the right side. This time interval is accompanied by the solar wind pressure increase ($\delta Psw = 9$ nPa at 19.2 UT), compressing the magnetospheric cavity ($\delta MP= -0.5$ $R_E$), accordingly, the SymH index registered an increase of $\sim$ 25 nT. However, the stations on the left panel show a positive bay type of signature with an amplitude of $\sim$ 50 nT. The magnetic measurements at Indian stations from Gulmarg to Tirunelveli and further crossing into the southern hemisphere up to $12^\circ$ GG latitude show positive variation of $\sim 50$ nT amplitude. This higher amplitude could be due to the additional effect of the positive bay associated with the substorm. The presence of Pi2 activity at  $\sim$ 19.5 UT suggests the expansion phase of the substorm. The plots shown in the right panel exhibit a decrease between two vertical dashed lines. The stations located north of Indian stations (NVS, IRT, and BMT) show strong negative excursion. The negative variation is strongest at NVS (GGLat= $54.8 ^\circ N$) of  $\sim 250$ nT and is observed at lower latitudes up to BMT located at $40.3^\circ$N with amplitude of $\sim$ 50 nT. Similarly in the southern hemisphere, LRM and GNG stations with GG latitudes of $22.2^\circ $S and $31.4^\circ$ S showed negative excursions of magnitudes 75 nT and  $\sim 100$ nT respectively. All five stations shown in the right panel are mid-latitude stations with geomagnetic latitude $< 50^\circ$ located near midnight, where positive H variation is expected. Even otherwise, due to the presence of solar wind pressure impulse, positive signatures at those stations are expected, but we observe a negative dip reaching up to $40^\circ$ N and $22^\circ$ S GG latitude. We further investigate this event using particle flux data and field-aligned currents (FACs), particularly in the time window between 19 to 20 UT, for identifying the equatorward boundary of auroral electrojet.

\section{Estimation of Equatorward Boundary of Auroral oval }

\subsection{Observations of particle flux by NOAA/POES satellites}

\begin{figure}[h!]
	\centering
	\includegraphics[width=15.0cm]{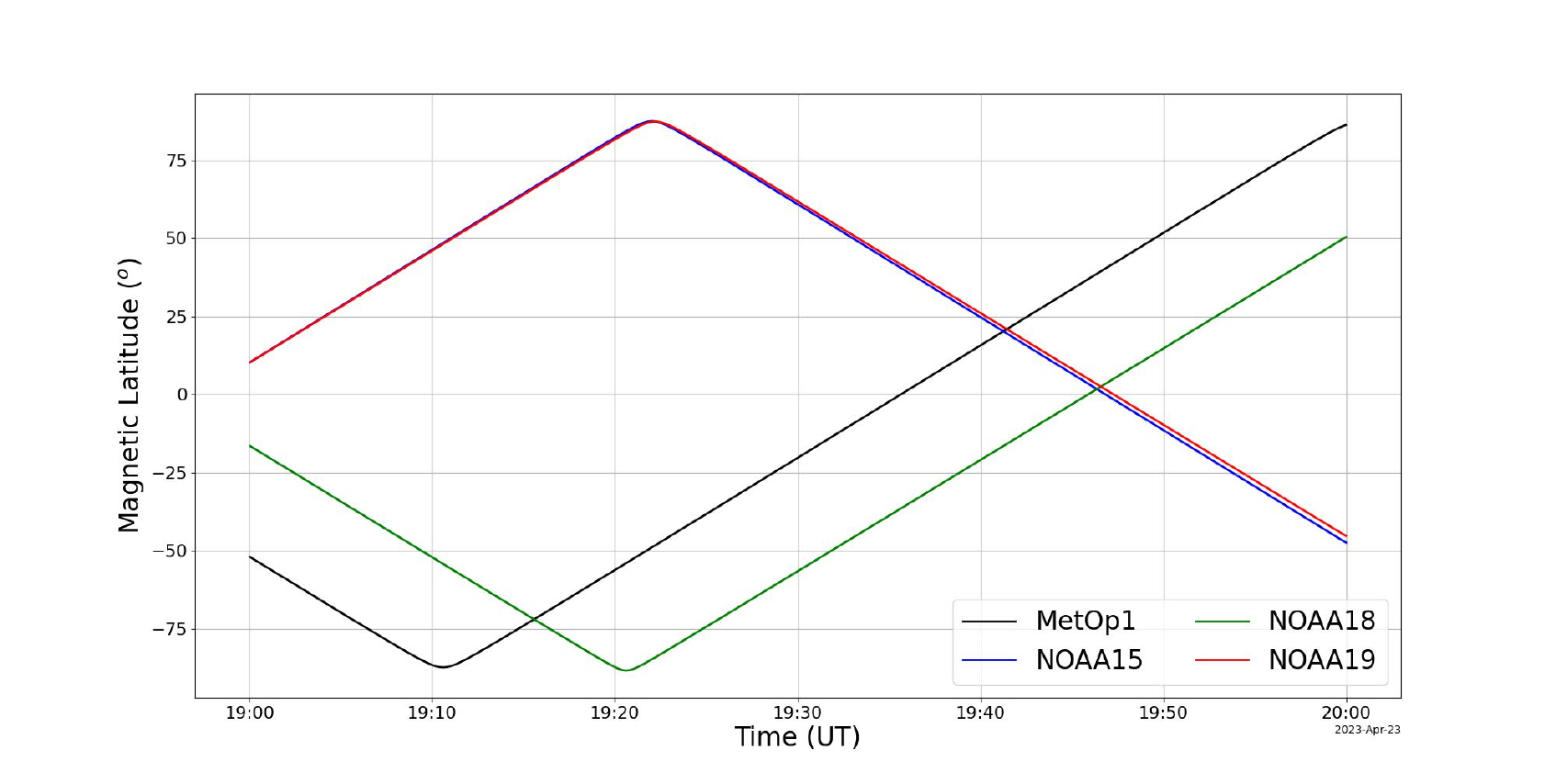}
	\caption{Trajectories (Magnetic latitude- Time frame) of various NOAA (N15, N18, N19) and Metop1 satellites during 19 to 20 UT on 23 April 2023.}
	\label{fig:lat_time}
\end{figure}
\begin{figure}[h!]
	\centering
	\includegraphics[width=15.0cm]{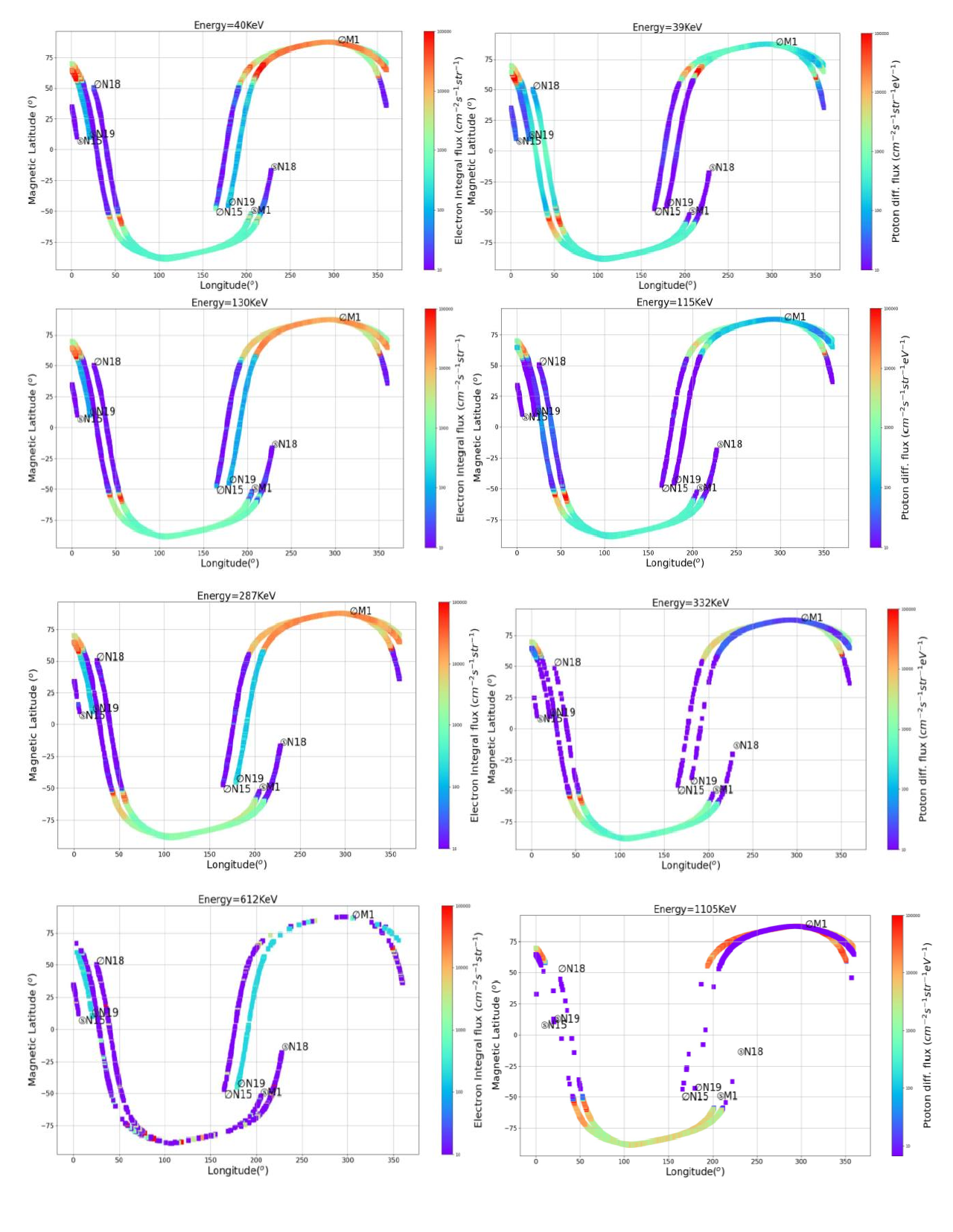}
	\caption{Electron flux (left) and proton flux (right) of various energies in Magnetic latitude- GG longitude frame. The color scale is the same in all the plots. The start and end of each satellite trajectory are indicated by  $\circledS$ and  $\varnothing$ symbols respectively.
}
	\label{fig:lat_time_flux}
\end{figure}

NOAA Polar-orbiting Operational Environmental Satellites (POES) are low earth orbiting satellites at $\sim$ 850 km altitude with nearly circular, sun-synchronous, polar orbits with an orbital period of $\sim$102 minutes. The Medium Energy Proton and Electron Detector (MEPED) instrument on board the POES satellites measures energetic protons and electrons $> 30$ keV in two orthogonal directions. Figure \ref{fig:lat_time} shows the trajectories of NOAA satellites (N15, N18, and N19) and Metop1 satellites in magnetic latitude (MLat)- time frame during 19 to 20 UT on 23 April 2023. Figure \ref{fig:lat_time_flux} displays the particle flux in the MLat- GG longitude frame of  Field-aligned/  $0^\circ$ directed (precipitating) electron (left) and proton fluxes (right) of different energies observed by N15, N18, N19, and Metop1, during 19-20 UT on 23 April. The satellites passing near $20^\circ$ longitude have a local time of around 20-21 hrs, while passes over $180^\circ$ longitude have $\sim$ 7-8 LT that can be considered as near dawn passes. The maximum precipitating flux is around $10^5$ (shown by the red color) and is located beyond $50^\circ$ MLat latitude. That means the equatorward auroral boundaries have not reached latitudes lower than $50^\circ$ MLat. In the northern hemisphere, the electrons with 40 keV energy indicate the equatorial boundary of precipitation as $54^\circ$N MLat, and proton flux of energy 39 keV shows this boundary at  $\sim 53^\circ$N MLat. It is noticed that lower energy protons (40 keV) have considerable flux (a few thousand) at low latitudes in N18 and M1 satellites, which traversed near midnight, while electrons showed minimum flux therein. The proton flux in N15 and N19 satellite passes traversing near dawn is minimal. Interestingly, N19 at $\sim$ 7.5 LT showed an electron flux of a few hundreds, but N15 at 6.5 LT showed an electron flux of a few tens. The electron flux recorded by N19 was always a few hundred at lower latitudes irrespective of energies and local time and is not consistent with nearby N15 observations. Therefore, we do not interpret N19 electron flux observations here. Near mid-night passes of N18 and M1 showed minimum electron flux at all energies, but a few thousands of protons of low energies at low latitudes.

\subsection{Low energy plasma flux using DMSP observations}

\begin{figure}[h!]
	\centering
	\includegraphics[width=15.0cm]{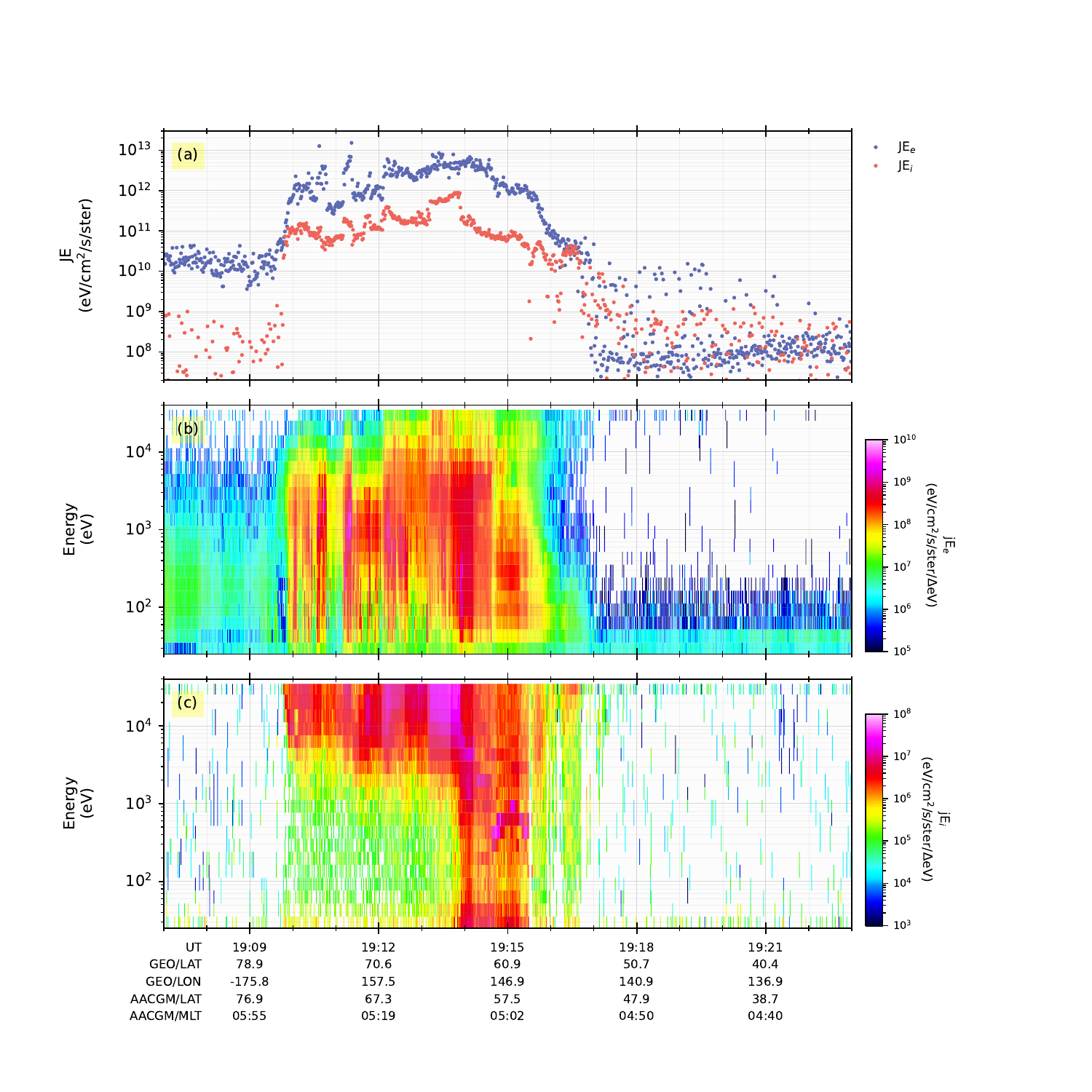}
	\caption{(a) Total particle flux in entire low energy band for electrons and ions; (b) Energy-time spectra for electrons; (c) Energy-time spectra for ions, during 19:07 to 19:23 UT on 23 April 2023. }
	\label{fig:dmsp_1}
\end{figure}

\begin{figure}[h!]
	\centering
	\includegraphics[width=15.0cm]{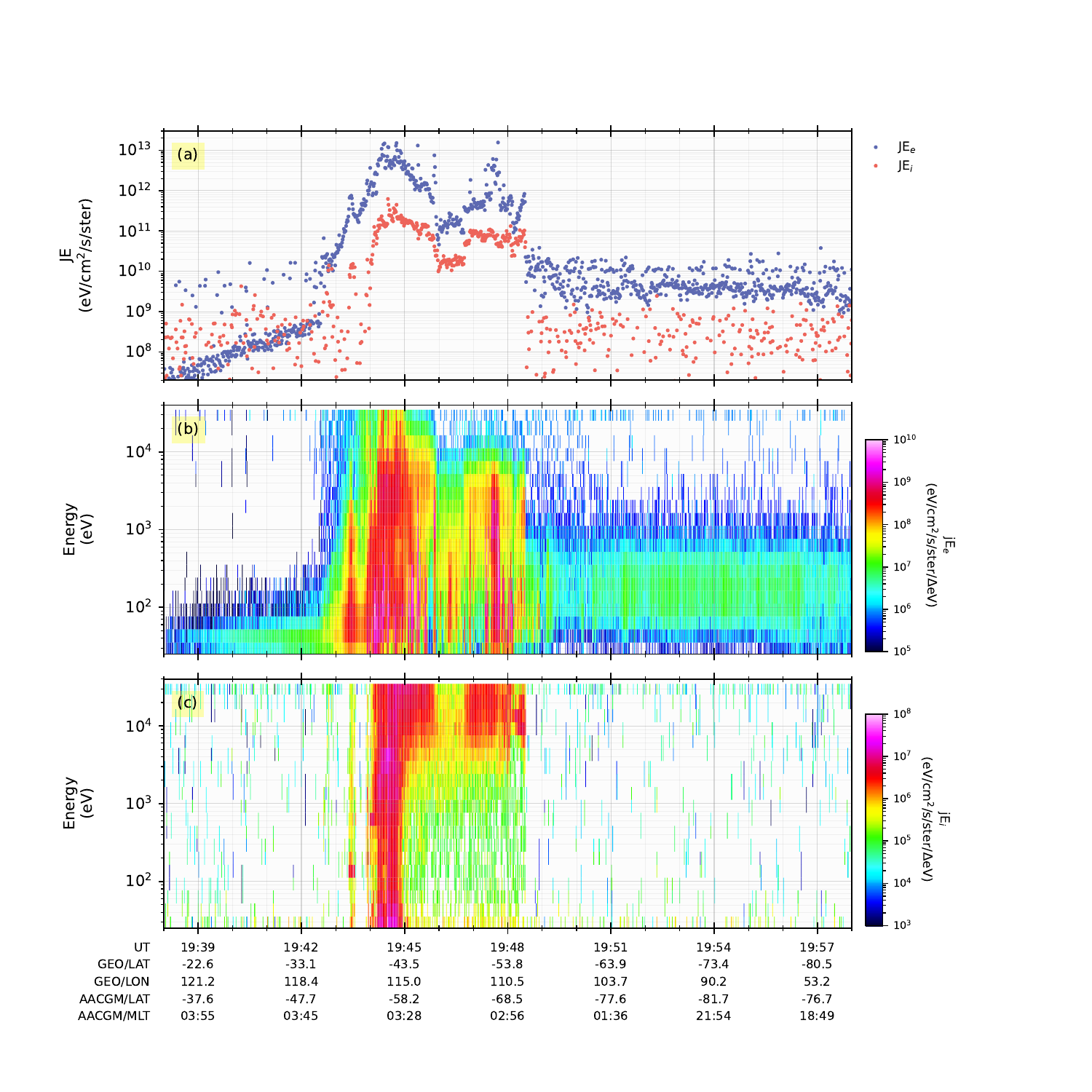}
	\caption{ (a) Total particle flux in entire low energy band for electrons and ions; (b) Energy-time spectra for electrons; (c) Energy-time spectra for ions, during 19:38 to 19:58 UT on 23 April 2023. }
	\label{fig:dmsp_2}
\end{figure}

Defense Meteorological Satellite Program (DMSP) satellites orbiting at $\sim$830 km altitude measure low energy (30 eV to 30 keV) electron and proton flux densities, thus providing the broadband plasma flux. \citet{shiokawa1997multievent} have shown the intensification of precipitating broadband electron (BBE) flux near the equatorward boundary of the particle precipitation region. Therefore, the BBE signature can be used to identify the equatorward boundary of the auroral activity. Figure \ref{fig:dmsp_1} and \ref{fig:dmsp_2}  show the DMSP-F18 low energy plasma flux observations during 19 to 20 UT in the northern and southern hemispheres respectively. It is dispensable to mention that the UT intervals in Figure \ref{fig:dmsp_1} and \ref{fig:dmsp_2} are different. The presence of BBE is evident in the plots. In Figure \ref{fig:dmsp_1}, the equatorward boundary associated with BBE is found to be at around $57.5^\circ$ GG latitude in the northern hemisphere,  which corresponds to $54.3^\circ$N CGM latitude. From Figure \ref{fig:dmsp_2}, the boundary is found to be at $- 38.3^\circ$ GG latitude ($53^\circ$S CGM) in the southern hemisphere. The low energy electrons (few tens of eV) are observed up to $54^\circ$N GG ($51^\circ$N CGM) and $35^\circ$S GG ($49.4^\circ$S CGM) latitudes. These low-energy electrons show peak flux at $\sim$ 60 to 100 eV. The low energy ions are observed up to $55^\circ$N GG ($52^\circ$N CGM) latitude.

\subsection{Auroral oval boundaries from Swarm mission}
\begin{figure}[h!]
	\centering
	\includegraphics[width=15.0cm]{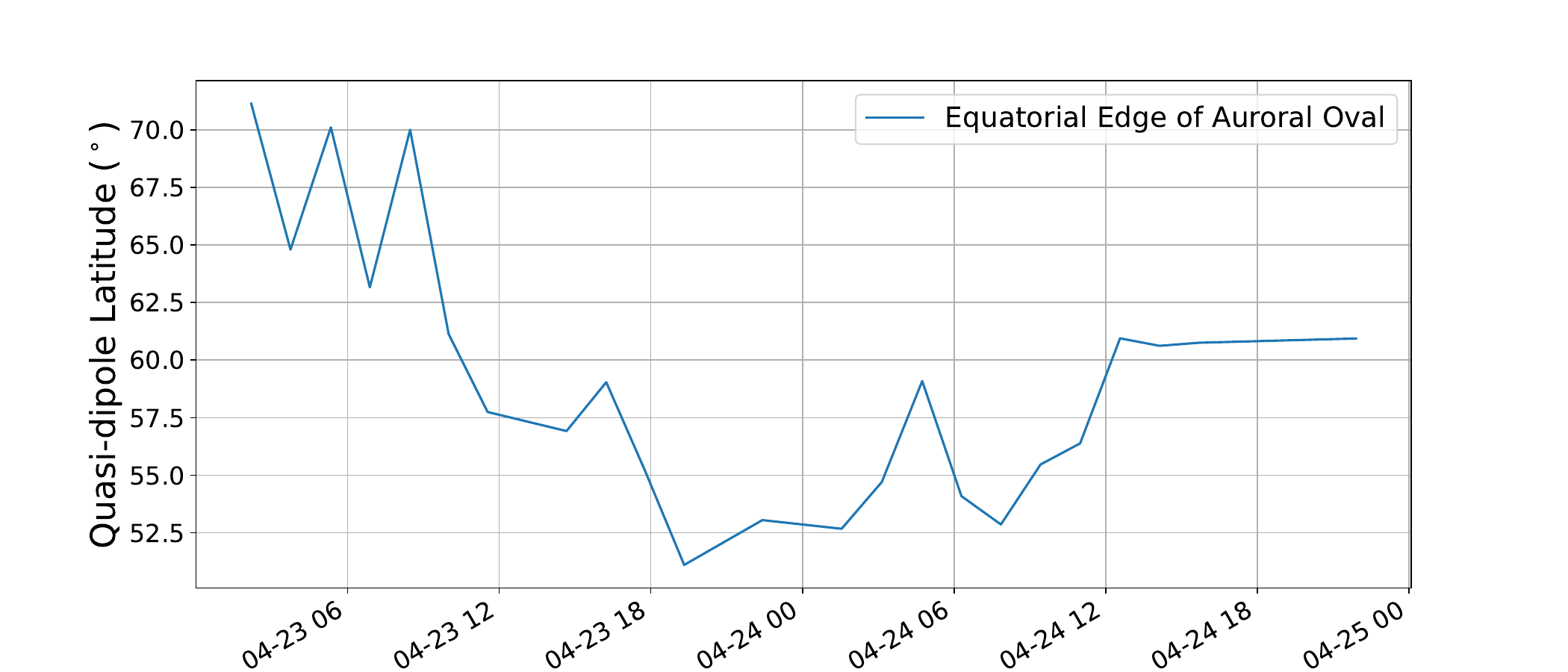}
	\caption{Swarm-based equatorward boundary in quasi-dipole latitudes in Northern hemisphere during 23-24 April 2023.
 }
	\label{fig:swarm_fac}
\end{figure}

The Swarm mission consisting of three polar low earth orbiting satellites provides the parameters related to auroral electrojet named AEBS (Auroral Electrojet and auroral Boundaries estimated from Swarm observations) products.  It estimates field-aligned currents (FACs) and derives the auroral oval boundaries based on them by merging electric fields through the Newell coupling function. The details of boundary determination can be found in \citet{xiong2014determining}. They found that the equatorward boundary has a linear dependence on Dst and AE indices, however, the equatorward expansion saturates toward high activity. Figure \ref{fig:swarm_fac}   shows the auroral oval boundaries derived from small and medium-scale FACs. It is seen from the plot that during 19-20 UT, the equatorward boundary is lowest at $\sim 51^\circ$ quasi-dipole latitude. This appears consistent with the observations from NOAA/POES and DMSP satellites.

\section{Discussion}

The red-colored emission was observed at Hanle, Ladakh near the northern horizon at $\sim 80^\circ$- $85^\circ$ from zenith (i.e $\sim 5^\circ$- $10^\circ$ elevation angle from the horizon ), which began to appear from $\sim 19.5$ UT till 22 UT on 23 April. This time corresponds to post-midnight (1.0 to 3.5 IST) time on 24 April. In fact, a red-colored diffuse aurora was seen on the same night from Xinjiang ( $44^\circ$ N GG latitude), in North-West China (\url{https://news.cgtn.com/news/2023-04-25/Stunning-auroras-amaze-skywatchers-in-Xinjiang-NW-China--1jhO0ZViIAU/index.html}).  Also there were reports of the aurora from Australia and New Zealand (\url{https://www.theguardian.com/world/gallery/2023/apr/25/aurora-australis-borealis-northern-southern-lights-auroras-across-the-world-after-solar-storm-pictures}; \url{https://www.theguardian.com/world/2023/apr/25/aurora-australis-new-zealand-sky-spectacular-light-show-solar-tsunami-southern-lights}). People in Europe and North America also witnessed the northern lights during the same event. However, due to the summertime sunlight in the northern hemisphere during April, spectacular auroras could not be noticed near the auroral oval. In the southern hemisphere of Antarctica, the red and green aurora was witnessed (Figure 1 and Supplementary Figure S2).
The aurora on the night side is usually associated with the substorm phenomenon comprising auroral electrojet currents, leaving a signature in the ground magnetic field measurements. Correspondingly, the auroral indices AL, and AE showed the occurrences of substorms during 16-22 UT, and the most intense ones (AL$< -1500$ nT) occurred during 19 to 22 UT. Night-time mid-latitude Pi2 oscillations, also represented by the Wp index \citep{nose2009new}, are directly related to the sudden development of substorm currents at the onset time leading to the substorm expansion phase. Since the reported event is a recent one, the Wp index is not available. However, we examined the one-second data from Indian observatories, which were on the night side. The figure presented in the supplementary material (Figure S1) shows the presence of Pi2 and hence confirms the substorm expansion phase after 19 UT on 23 April 2023, which also coincides with the auroral observations from Hanle. 

The statistical studies have reported that the super substorms mainly take place during the sheath plasma compression and magnetic clouds (MCs) of ICMEs  \citep{hajra2016supersubstorms,despirak2019supersubstorms}. The magnetospheric cavity compression generating waves can trigger the substorms, and it is now well established that the whistler mode chorus waves efficiently transfer the energy from an abundant population of low-energy electrons (10-100 keV) in the radiation belt to higher energy electrons, thus they are responsible for the acceleration as well as precipitation of radiation belt electrons \citep{horne2005wave,meredith2001substorm, thorne2005timescale,shprits2009evolution,shprits2008review}. On 23 April 2023, during the intense substorm activity, the solar wind pressure was quite high, compressing the magnetosphere to 6 $R_E$ (Figure \ref{fig:IP} ). The GOES 16 and 17 satellites were outside the magnetosphere \citep{ghag2024quasi}. Such compression during the passage of ICME sheath along with the prolonged southward IMF might have triggered such an intense substorm on 23 April 2023. Moreover, solar wind pressure impulse compresses the entire magnetospheric cavity, moving the inner boundary of the tail current sheet closer to the Earth, which can result in lower latitude precipitation.

The magnetic signature of westward auroral electrojet is due to the substorm current wedge (SCW), which comprises the cross-tail currents, FACs, and ionospheric Hall currents \citep{mcpherron1973satellite}. SCW produces a distinctive magnetic field pattern of the positive bay in the H component at the mid-latitude stations located within 21-02 LT hours \citep{clauer1974mapping}. We examined the magnetic field variations at the stations from Indian-Asian longitudes ( $70^\circ - 115^\circ $ E longitude) on the night side. The storm time magnetic field signatures at mid to low latitudes are mainly dominated by the ring current monitored by the SymH index, and hence the positive bays in the H variations associated with the SCW system are not easily discernible. The positive H variation between 19.2 to 19.8 UT exhibits bay-type variation, although it was also accompanied by the solar wind dynamic pressure impulse that produced positive variations on the ground. As mentioned above, this timing also coincided with the appearance of red-colored aurora at Hanle, intense AL index, and Pi2 occurrence. The amplitude of positive increase due to the solar wind pressure impulse is $\sim 25$ nT in the SymH index, whereas the amplitude of H variation on the night side mid-latitude stations is $\sim 50$ nT, which could be due to the additional contribution of SCW-related positive bay. Figure \ref{fig:mag_var} indicates that not all mid-latitude stations show this positive variation. A rapid and deep depression of 250 nT magnitude in the H component is observed at the NVS ( $51.75^\circ$ N CGM latitude) station. This depression was seen in both hemispheres at 5 mid-latitude stations located near midnight, reaching up to  $36^\circ$N and  $32^\circ$ S CGM latitudes. This indicates the presence of the signature of substorm at mid-latitudes, in the northern and southern hemispheres, despite the expected positive variation due to the positive bay and solar wind pressure impulse. The high-latitude stations located north of NVS viz. AMD ( $66.08^\circ$ N CGM), DIK ( $69.94^\circ$ N CGM), and BRN ( $74.78^\circ$N CGM) show negative depression in H of magnitude 1000 nT, 1500 nT, and 500 nT respectively, indicating a strong westward electrojet/SCW there. Now the question is: Does the negative depression at lower latitudes indicate the expansion of the equatorward boundary of the auroral oval? The auroral boundaries observed from the particle flux data in NOAA/POES and DMSP satellites indicate the expansion of the equatorward boundary is up to  $\sim 51^\circ$- $54^\circ$N CGM. This boundary is also consistent with the Swarm-AEBS products of auroral boundaries derived from FACs and solar wind-magnetosphere coupling function. Thus, it is clear from the particle flux and electric current observations that the boundaries have not shifted below $50^\circ$.  \citet{xiong2014determining} have shown that the auroral boundaries and the size of the auroral oval have significant dependence on the magnetic activity and local time sector, however, the midnight equatorward boundary saturates at higher activity. According to the estimates calculated by \citet{blake2021estimating} for a geomagnetic storm with minimum $Dst=-200$ nT, the maximum extent of the auroral equatorward boundary (MEAEB) would be above 50° GM latitude. \citet{landry2019empirical} found that the equatorward boundary of the diffuse aurora is best described by a weighted average of the AE index, rather than the Dst or Kp indices. Even with this consideration, the MEAEB is found to be $> 50^\circ$. Though the equatorward boundary is more equatorward during equinoxes than solstices, this difference in latitude is not very significant. Thus, the models do not estimate the equatorward boundary to expand below  $50^\circ$ latitude during the substorms with AL $\sim -1500$ nT, which is also evident from the particle precipitation data of NOAA POES satellites as mentioned above. Then why the negative H depression and the aurora are observed at low to mid-latitudes?
The westward current giving negative H variation at lower latitudes could be due to the ring current and magnetotail current. But the signature discussed in the present paper (between two vertical lines in Figure \ref{fig:mag_var}) is accompanied by the solar wind pressure impulse, which shows clear positive variation at other lower latitudes and is also expected to have a similar nature at stations like BMT, LRM, unlike the actual observations. In the absence of overhead ionospheric currents at nightside mid-latitudes close to the auroral boundary, the strong auroral ionospheric currents influence the magnetic field signatures. The westward auroral electrojet currents flowing in the ionosphere of the auroral oval create negative H variation just below the oval, as well as at lower latitudes than that of the oval. When the H depression has a magnitude of $\sim 1500$ nT in the auroral oval, the lower latitudes still capture the negative H variation of magnitude $\sim 50$ nT. Accordingly, the Z variation at these lower latitudes in the northern hemisphere is upward (negative), and downward (positive) in the southern hemisphere.

\begin{figure}[h!]
	\centering
	\includegraphics[width=8.0cm]{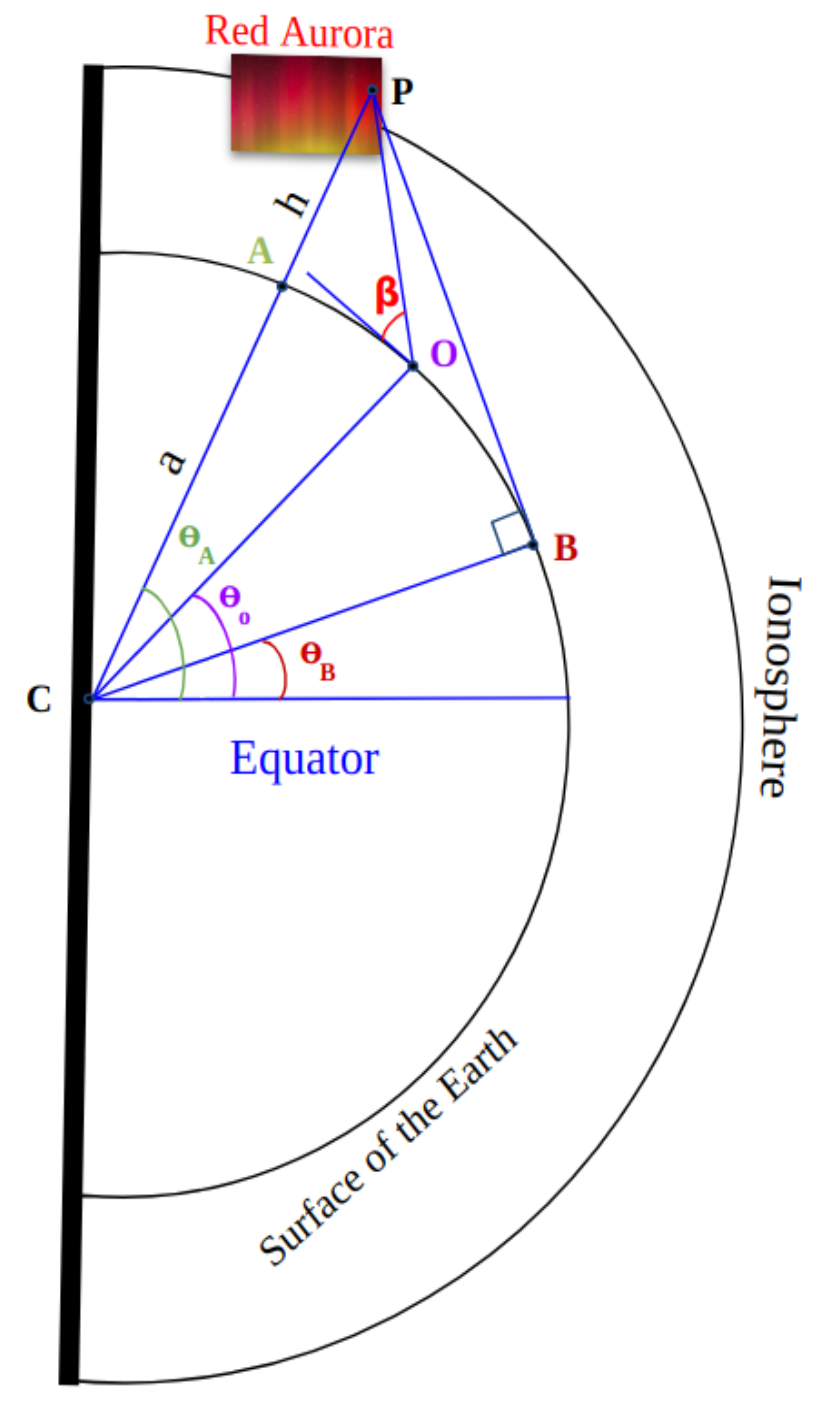}
	\caption{Sketch showing the geometry of Auroral visibility at observer latitude($\theta_o$).  C is the center of the Earth; a is the radius of the Earth: 6370 km; P is the point of auroral precipitation at height ‘h’ km at geographic latitude $\theta_A$, and A is the point on the surface of the Earth. B is the tangent point at the surface of the Earth line originating from point P, and $\theta_A$ is the latitude of that point. $\beta$ is the maximum elevation angle of the aurora seen by the observer. The sketch is not to scale.
 }
	\label{fig:sketch}
\end{figure}

Many researchers have attributed the source of diffused red aurora to the hot electrons coming from the central plasma sheet (CPS)  \citep{ni2011resonant,ni2011chorus,nishimura2022interaction,horne2000electron,thorne2010scattering,liang2011fast,zhang2014excitation}. The DMSP observations of low energy plasma (10 eV to 30 keV) indicate the precipitation of broadband electron (BBE) flux at the equatorward boundary of the auroral precipitation, which originates from the inner part of the plasma sheet and is attributed as a possible cause of red aurora observed at mid-latitudes  \citep{shiokawa1997multievent}. The low energy electrons of BBE interact with the Oxygen atoms at higher altitudes $ \sim 400$ km and cause red color emissions (630.0 nm). Whereas higher energy electrons penetrate deeper into the ionosphere causing green color aurora (577.7 nm) on interaction with Oxygen at $\sim 100$ to 300 km height. The observed aurora at Hanle was not overhead, rather it was seen near the northern horizon. The observed red color appeared diffused without a sharp boundary. Using the background star positions in the sky and simulated night sky by Stellarium Planetarium software, the elevation angle ($\beta$) of the observed red light is estimated as 5- $10^\circ$.  Figure \ref{fig:sketch} shows the schematic geometry of the auroral visibility at observer. The auroral emissions occurring over point ‘A’ at a height of 'h' from the surface of the Earth.  Angle $\theta_A$ is the latitude of A and $\theta_o$ is the latitude of the observer location 'O'.  If the angle of elevation ($\beta$) of auroral light at point 'O' is known, then the height of auroral emissions over point 'A' can be calculated as,

\begin{equation}
 h = \frac{a . cos (\beta)}{cos (\beta + (\theta_A - \theta_o))} - a
\end{equation}



Where $'a'$ is the radius of Earth = 6370 km.  
Point 'B' is the location where auroral emissions will be at the horizon, i.e. $\beta$= 0, so it can be considered as the lowest latitudinal boundary of the auroral visibility. For the red aurora occurring over point A at 500 km altitude, the aurora will be seen within a periphery of the radius of angle $ \sim22^\circ$. The scattering effects due to dense atmosphere are found to be less than half a degree.
From the previous sections, we find that for the 23 April 2023 event, the low energy electrons precipitate up to $\theta_A$ = $54^\circ$N GG latitude. For $\beta$= $10^\circ$ at Hanle location ($\theta_o$= $33^\circ$ GG), h is found to be 948.5 km. For $\beta$= $7^\circ$ and $5^\circ$, h estimates to 791 km and 690 km respectively. The red aurora observed from China ($44^\circ$ N GG) at elevation angles of $\sim 30^\circ$ gives an altitude of around 830 km, which is consistent with the estimates obtained from Hanle observations. Thus, the red color emissions occurring at $54^\circ$ N GG location at the height $\sim$ 700- 950 km, would be responsible for the observed aurora at Hanle.  As per the previous reports, the 630 nm emissions can be present up to 600 to 650 km in height. However, the present calculations indicate little higher altitudes. Owing to the precipitation of low energy electrons of a few tens to hundreds of eV as seen in Figure 7, the 630 nm emission at these higher altitudes is possible, The green aurora produced at $\sim 100$ km can reach above $54 -10= 44^\circ$ GG latitude, and cannot be visible from the lower latitudes. It is possible that green, red, and blue auroras were present in the auroral oval (Figure \ref{fig:sup2} from Antarctica). Also, a combination of various colors creates more shades of color like pink, purple, cyan, etc. However, due to higher altitudes, the red color is prominently seen from the lower latitudes. This is why the low to mid-latitude aurora is normally red in color \citep{tinsley1986low}. 
 
 Also interestingly, the NOAA/POES observations show that although the integral flux of low energy (40 keV) protons is maximum in the auroral oval, a few thousand (two orders smaller than the maximum) of protons precipitated at lower latitudes in the midnight sector, whereas the precipitating electron flux is minimum therein. The solar wind pressure impulse-generated waves such as Pi2, EMIC could be responsible for the scattering of plasma particles from the ring current, which follow the magnetic field lines. However, the presence of particles close to the equator is intriguing. The auroral images shown in Figure  \ref{fig:sup2} indicate STEVEs as well as SAR arc-kind features. The scattered protons could be responsible for the SAR arc, as it is produced by the ionospheric heating due to the precipitation of protons.

\section{Conclusions}

On the night of 23-24 April 2023, Hanle observatory in Ladakh, India observed a red color light glow in the north direction with the all-sky camera having $180^\circ$ field of view, which was claimed as an aurora. This is the first observational imaging evidence of low-latitude aurora in India. It is extremely rare to observe aurora at such low latitudes during a moderate geomagnetic storm. However, this modern-day observation of low latitude aurora in conjunction with in situ multi-spacecraft particle measurements, and ground, and satellite-based magnetic field measurements provide some meaningful understanding of it. Following are the concluding remarks:

\begin{itemize}
    \item Intense substorms occurred during the period of red aurora observations from Ladakh.
    \item In situ, plasma particle data, and Swarm mission AEBS products show that the equatorward boundary of the auroral precipitation has moved towards lower latitudes compared to other active periods (up to $54^\circ$ GG latitude), although it has not reached Indian latitudes.
    \item The solar wind pressure impulse compressed the entire magnetospheric cavity, which could have moved the inner boundary of the tail current sheet closer to the Earth. 
    \item Occurrence of BBE event is found. However, the low energy electrons ($< 100$ eV) precipitating at $\sim$ $54^\circ$ GG latitude are found to be responsible for producing red light emissions at higher altitudes ($\sim$ 700 - 950 km). Simple calculations demonstrate that this aurora can be within the line of sight from low latitudes such as Hanle. 
    \item Thus, the red aurora that is observed from the lower latitudes is unequivocally the result of two factors. Firstly, the red emissions that occur at higher altitudes in the auroral oval, and secondly, a slight expansion of the oval towards the equator. It is important to note that low-latitude aurora is not caused by the auroral oval's equatorial boundary reaching low latitudes.
    \item Negative H variation at nighttime mid-latitudes below the equatorward boundary of auroral precipitation is observed, which could be due to the westward auroral electrojet currents flowing in the ionosphere of the auroral oval.
    \item SAR arc and STEVE-type features were observed from Antarctica during this event.
    \item The observation of a considerable amount of proton flux at low latitudes could be due to the scattering of ring current particles, which might be responsible for the SAR arc associated with the auroral signatures at Hanle, India.

\end{itemize}

\acknowledgments
The authors acknowledge the IIA camera observation records at Hanle, India, and NCPOR’s All Sky Camera observations from Maitri station, Antarctica. We acknowledge the INTERMAGNET and IIG magnetic observatory network for providing ground geomagnetic field data. We thank WDC-Kyoto for providing geomagnetic indices, and the Space Physics Data Facility/Goddard Space Flight Center OMNIWeb Interface (https://omniweb.gsfc.nasa.gov/) for providing solar wind and IMF data. The authors acknowledge the DMSP, NOAA/POES, and Swarm mission for the data used in the present study. The solar wind parameters, interplanetary magnetic field, and geomagnetic indices used in this study are obtained from CDAWEB (\url{https://cdaweb.gsfc.nasa.gov/}) and SUPERMAG (\url{http://supermag.jhuapl.edu/}).

\clearpage
\section*{Supplementary Informtion}

\renewcommand{\thefigure}{S1}
\begin{figure}[h!]
	\centering
	\includegraphics[width=12.0cm]{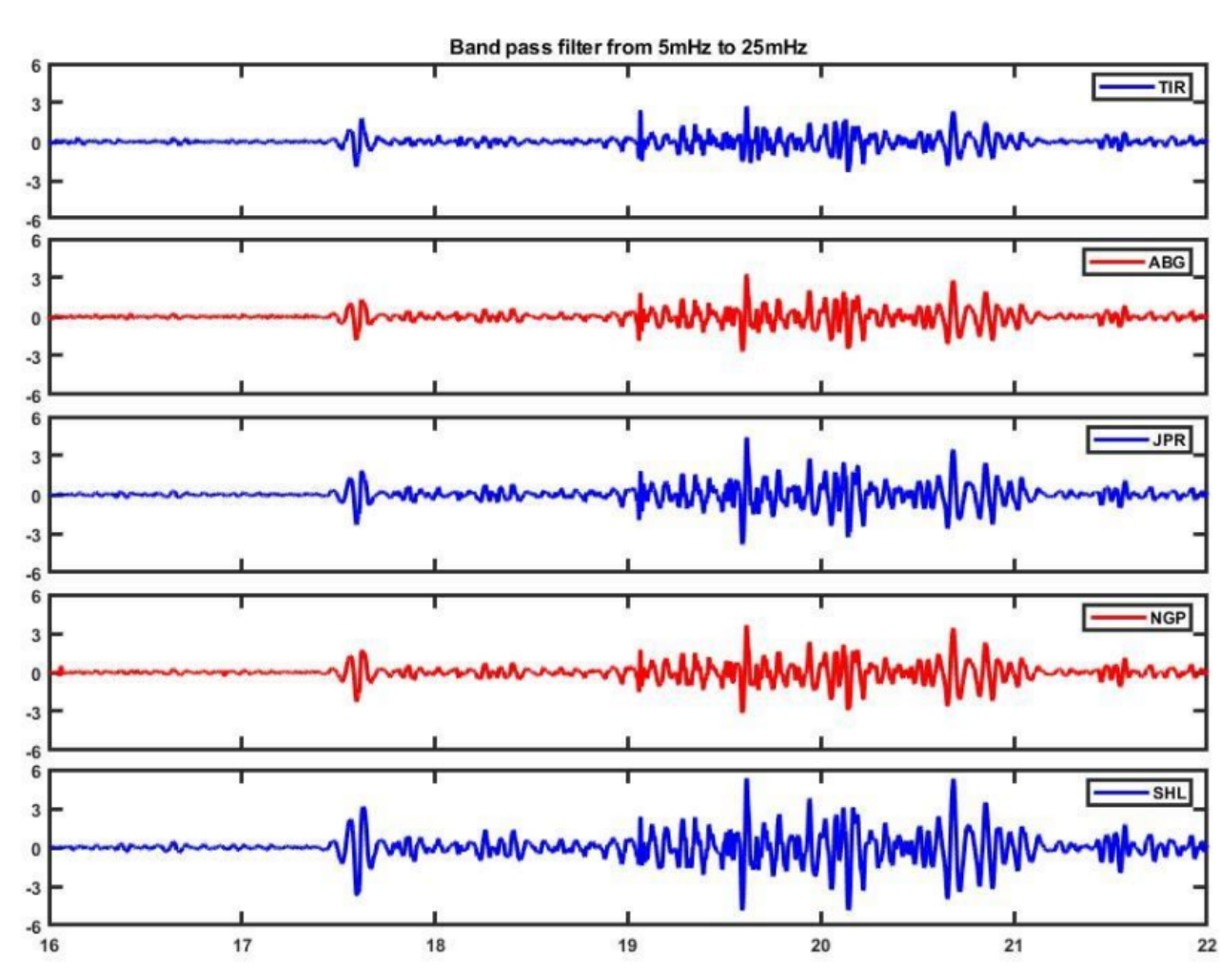}
	\caption{Pi2 pulsations observed at low latitude night stations- band pass between 5-25 mHz.
 }
	\label{fig:sup1}
\end{figure}

\renewcommand{\thefigure}{S2}
\begin{figure}[h!]
	\centering
	\includegraphics[width=15.0cm]{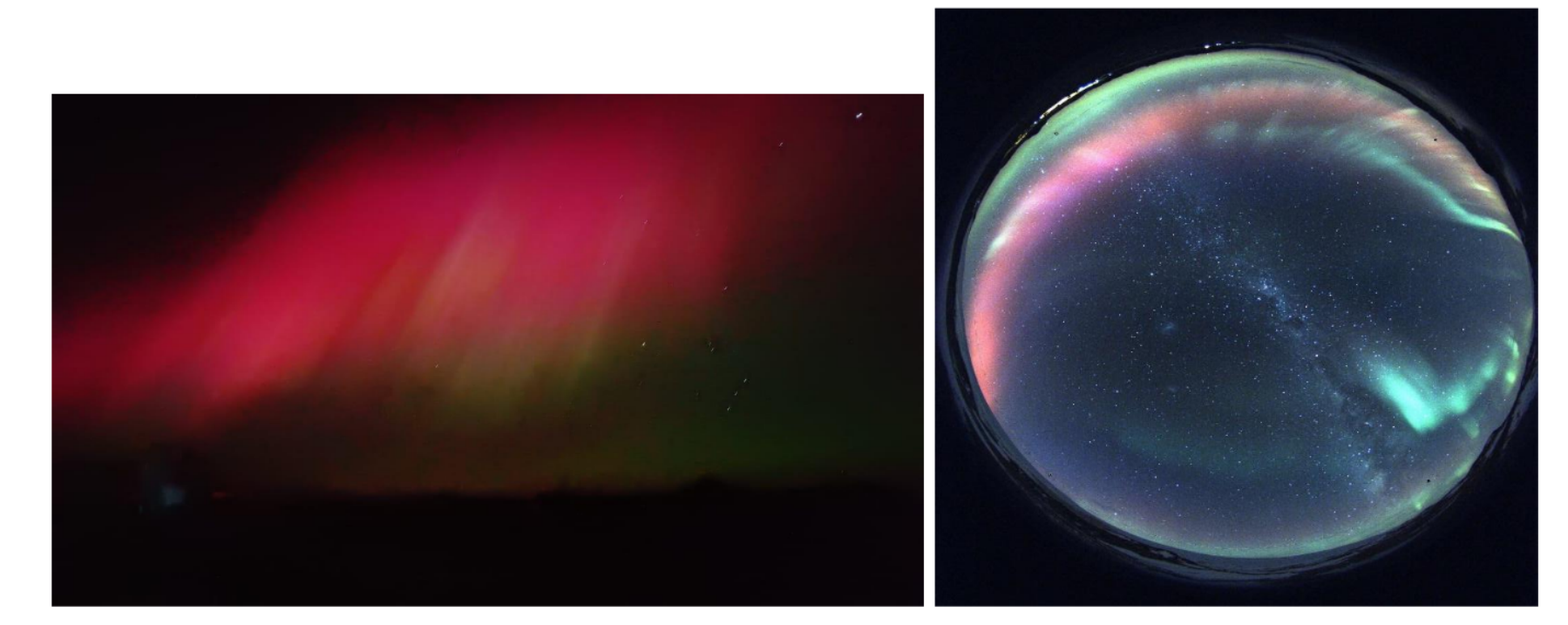}
	\caption{Aurora seen from Indian Antarctic station, Maitri on 23 April 2023 (a) photograph at 22:01 UT, All Sky camera (b) at 19:22 UT.
 }
	\label{fig:sup2}
\end{figure}

\begin{table}[htbp]
\centering
\label{table:station-info}
\begin{tabular}{@{}ccccccc@{}}
\toprule
\textbf{S.No.} & \textbf{Station Name} & \textbf{IAGA CODE} & \textbf{GLAT} & \textbf{GLONG} & \textbf{GMLAT} & \textbf{CGMLAT} \\ \midrule
1 & Gingin & GNG & -31.36 & 115.72 & -43.09 & -42.67 \\
2 & Learmonth & LRM & -22.22 & 114.10 & -32.14 & -32.23 \\
3 & Cocos-Keeling Islands & CKI & -12.19 & 96.83 & -22.26 & NA \\  
4 & GAN & GAN & -0.70 & 73.15 & -7.77 & NA \\
5 & Tirunelveli & TIR & 8.70 & 77.00 & 0.03 & NA \\ 
6 & Pondicherry & PON & 11.92 & 79.92 & 3.07 & NA \\
7 & Dalat & DLT & 11.95 & 108.48 & 1.96 &NA \\ 
8 & Chutuppal & CPL & 17.29 & 78.92 & 10.24 & NA \\ 
9 & Hyderabad & HYB & 17.42 & 78.55 & 8.65 & NA \\ 
10 & Visakhapatnam & VSK & 17.68 & 83.32 & 8.56 & NA \\ 
11 & Alibag & ABG & 18.64 & 72.87 & 10.37 & NA \\ 
12 & Phuthuy & PHU & 21.00 & 106.00 & 11.05 & 15.37 \\
13 & Nagpur & NGP & 21.15 & 79.10 & 12.33 & 15.79 \\
14 & Rajkot & RKT & 22.30 & 70.93 & 14.21 & 17.28 \\
15 & Sillchar & SIL & 24.93 & 92.82 & 15.27 & 19.73 \\
16 & Shillong & SHI & 25.55 & 91.88 & 15.95 & 20.42 \\
17 & Jaipur & JAI & 26.92 & 75.80 & 18.34 & 22.22 \\
18 & Gulmarg & GUL & 34.02 & 74.42 & 25.60 & 29.98 \\
19 & Beijing & BMT & 40.30 & 116.20 & 30.41 & 35.56 \\
20 & Irkutsk & IRT & 52.27 & 104.50 & 43.05 & 48.56 \\
21 & Novosibirsk & NVS & 54.80 & 83.20 & 45.80 & 51.75 \\ \bottomrule
\end{tabular}
\caption{Geomagnetic observatories information: The CGM latitudes obtained from \url{https://omniweb.gsfc.nasa.gov/vitmo/cgm.html} and are computed only for the geographic latitude greater than $20^\circ$. Quasi-dipole latitudes are taken from \url{http://www.geomag.bgs.ac.uk/data_service/models_compass/coord_calc.html} }
\end{table}

\clearpage
\bibliography{ref}
\bibliographystyle{aasjournal}

\end{document}